\documentclass[%
 aip,
 amsmath,amssymb,
 reprint,%
]{revtex4-1}

\draft 

\usepackage{xcolor}
\usepackage{soul}

\usepackage{graphicx}
\usepackage{dcolumn}
\usepackage{bm}
\usepackage{subcaption}
\usepackage[utf8]{inputenc}
\usepackage[T1]{fontenc}
\usepackage{mathptmx}
\usepackage{etoolbox}
\usepackage{physics}

\usepackage[font=small,labelfont=bf,
   justification=raggedright,
   format=plain,
   singlelinecheck=off]{caption}

\makeatletter
\def\@email#1#2{%
 \endgroup
 \patchcmd{\titleblock@produce}
  {\frontmatter@RRAPformat}
  {\frontmatter@RRAPformat{\produce@RRAP{*#1\href{mailto:#2}{#2}}}\frontmatter@RRAPformat}
  {}{}
}%
\makeatother
\begin{document}

\preprint{AIP/123-QED}

\title[Design and characterization of a cryogenic vacuum chamber for ion trapping experiments]{Design and Characterization of a Cryogenic Vacuum Chamber for Ion Trapping Experiments}
\author{D. M. Hartsell}
 \affiliation{Quantum Systems Division, CIPHER Laboratory, Georgia Tech Research Institute, Atlanta, GA 30318, USA}
 \email{darian.hartsell@gtri.gatech.edu}
\author{J. M. Gray}
 \affiliation{Quantum Systems Division, CIPHER Laboratory, Georgia Tech Research Institute, Atlanta, GA 30318, USA}
\author{C. M. Shappert}
 \affiliation{Quantum Systems Division, CIPHER Laboratory, Georgia Tech Research Institute, Atlanta, GA 30318, USA}
\author{N. L. Gostin}
 \affiliation{Quantum Systems Division, CIPHER Laboratory, Georgia Tech Research Institute, Atlanta, GA 30318, USA}
\author{R. A. McGill}
 \affiliation{Quantum Systems Division, CIPHER Laboratory, Georgia Tech Research Institute, Atlanta, GA 30318, USA}
\author{H. N. Tinkey}
 \affiliation{Quantum Systems Division, CIPHER Laboratory, Georgia Tech Research Institute, Atlanta, GA 30318, USA}
\author{C. R. Clark}
 \affiliation{Quantum Systems Division, CIPHER Laboratory, Georgia Tech Research Institute, Atlanta, GA 30318, USA}
\author{K. R. Brown}
 \affiliation{Quantum Systems Division, CIPHER Laboratory, Georgia Tech Research Institute, Atlanta, GA 30318, USA}

\date{1 October 2025}

\begin{abstract}
We present the design and characterization of a cryogenic vacuum chamber incorporating mechanical isolation from vibrations, a high numerical-aperture in-vacuum imaging objective, in-vacuum magnetic shielding, and an antenna for global radio-frequency manipulation of trapped ions. The cold shield near 4~K is mechanically referenced to an underlying optical table via thermally insulating supports and exhibits root-mean-square vibrations less than 7.61(4)~nm. Using the in-vacuum objective, we can detect 397~nm photons from a trapped $^{40}\mathrm{Ca}^{+}$ ion with 1.77\% efficiency and achieve 99.9963(4)\% single-shot state-detection fidelity in 50~$\mu$s. To characterize the efficacy of the magnetic shields, we perform Ramsey experiments on the ground state qubit and obtain a coherence time of 24(2)~ms, which extends to 0.25(1)~s with a single spin-echo pulse. XY4 and XY32 dynamical decoupling sequences driven via the radio-frequency antenna extend the coherence to 0.72(2)~s and 0.81(3)~s, respectively.

\end{abstract}

\maketitle


Trapped ions represent a widely used platform for quantum computing experiments. Fault-tolerant fidelities have been demonstrated for trapped-ion state preparation and readout\cite{Myerson_detect,sotirova2024,An2022}, single-qubit gates\cite{Brown_2011,smith2024}, and two-qubit gates\cite{Benhelm_2008, Clark_2021, oxford_2qubit, racetrack}. These operations have been combined into more complex algorithms within an architecture in which ions are shuttled among various locations between gate operations \cite{Pino_2021, racetrack, Srinivas_2021}. Preserving qubit coherence during long quantum algorithms requires isolation of the ion trap from external sources of noise. Mechanical cold-head vibrations can lead to gate amplitude errors when the ion vibrates with respect to the control optics. Ambient magnetic field fluctuations lead to dephasing of magnetic field sensitive states. Achieving the highest possible detection fidelity requires high-efficiency light collection and detection, often constrained by the geometry of the ion trap and vacuum chamber. Here we present a cryogenic vacuum chamber designed with vibration isolation, in-vacuum magnetic shielding, a movable in-vacuum high numerical aperture (NA) objective, and an in-vacuum radio-frequency (rf) antenna for driving magnetic-dipole transitions in trapped ions confined to a surface-electrode trap. We describe a vibrometer characterization of the trap vibrations, an analysis of the ion-fluorescence collection efficiency and state-detection fidelity, and measurements of the coherence time of a magnetically sensitive qubit.

\begin{figure}[!h]
\includegraphics[width=0.48\textwidth]{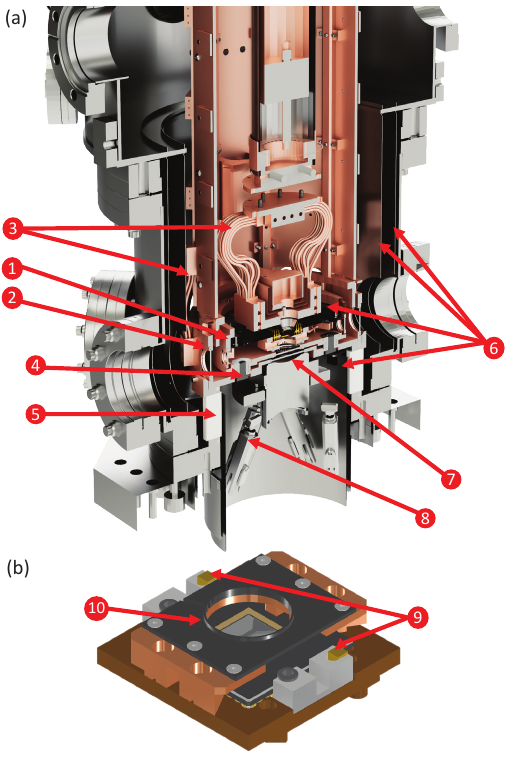}
\caption{\label{fig:chamber} \textbf{(a)} 3D-model cutaway illustrating the cryogenic vacuum chamber for ion trapping experiments. The cold \textbf{(1)} and intermediate \textbf{(2)} shields are thermally referenced to the cryocooler 4 and 50~K stages via flexible copper straps \textbf{(3)}. Mechanically, the chamber is referenced to the optical table via thermally insulating Vespel \textbf{(4)} and Macor \textbf{(5)} posts. Three layers of magnetic shielding material \textbf{(6)} are installed within the vacuum chamber. The imaging objective \textbf{(7)} is positioned on a piezo hexapod \textbf{(8)} at a working distance of 20.5~mm. \textbf{(b)} Enlarged view of the ion trap. Within the cold shield, two small magnets \textbf{(9)} establish the bias magnetic field, and a resonant coil \textbf{(10)} is placed above the trap to provide radio-frequency magnetic fields.}
\end{figure}

The vacuum chamber is pumped to $10^{-5}$~Torr using a turbo pump, then an ion pump and the cryostat are both turned on. Once cold, the chamber achieves a base pressure of approximately $10^{-11}$~Torr, as measured at the vacuum feedthroughs near the top of Figure ~\ref{fig:chamber}(a). The chamber is cooled using a two-stage Gifford-McMahon cryocooler (SHI Cryogenics RDK-415D2), powered by a helium compressor (Sumitomo F-70L), with 45~W of cooling power at the 50~K stage and 1.5~W at the 4.2~K stage. The expansion and compression cycle of the Gifford-McMahon cryocooler can create acoustic vibrations of greater than 20~$\mu$m\cite{Tomaru_2004} that can couple to mechanical resonances in the chamber and displace the ion trap relative to external optics. The mechanical coupling between the cold head and chamber can be greatly suppressed by suspending the cold head on a separate mechanical support and employing helium to conduct heat between the chamber and cold head in a gas exchange space; flexible bellows seal gas inside the exchange space with reduced vibration propagation to the chamber\cite{Dubielzig_2021}. The Gifford-McMahon cryocooler and closed cycle exchange gas system are part of an ultra-low vibration assembly supplied by Lake Shore Cryotronics (SHI-4XG-UHV-15). Even with these isolation mechanisms in place, vibrations from the cold head can induce relative displacements up to 50-100~nm root-mean-square (RMS), resulting in fluctuations in optical phase and intensity at the ion trap that contribute to considerable errors in laser-based gates\cite{pagano_2018}. These systems also experience slow drifts in position due to the support structure's long lever arm in combination with slow variations in temperature within the chamber. Newly designed cryogenic systems further suppress mechanical oscillations by thermally anchoring the ion trap to the 4~K stage via flexible copper braids\cite{heim, pendulum}. Vibrations of less than 20 nm RMS were achieved by utilizing a 1.7~m long inertial pendulum to decouple the ion trap from the cryocooler in a separate mechanical room\cite{pendulum}. We employ similar thermal anchoring to achieve low vibration levels but with a more compact design in which the ion trap is mechanically referenced to a floating optical table via short, thermally insulating support posts.

\begin{figure}[!ht]
    \includegraphics[width=0.48\textwidth]{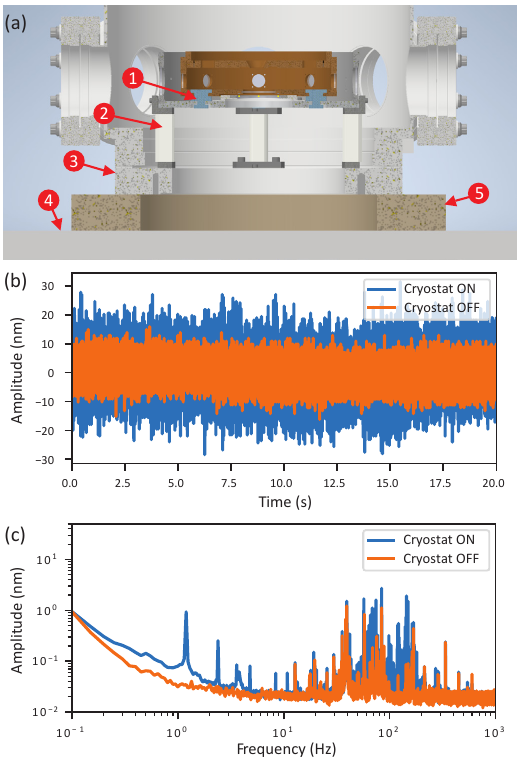}
\caption{\label{fig:vibration} \textbf{(a)} Schematic of vibration isolation components in the cryostat. Threaded Vespel bolts and posts \textbf{(1)} secure the cold shield to the intermediate shield. The intermediate shield is then bolted to Macor posts \textbf{(2)} which are bolted to the vacuum chamber \textbf{(3)}. The vacuum chamber is then bolted to the optical table \textbf{(4)} via a mounting plate \textbf{(5)}. This assembly mechanically references the ion trap to the optical table where addressing beams are mounted. \textbf{(b)} Time series vibration measurement of the cold shield. \textbf{(c)} Frequency components of vibration measurement in (b) with resolution bandwidth of 0.05~Hz.}
\end{figure}

A diagram of the cryogenic vacuum chamber is shown in Figure ~\ref{fig:chamber}(a); the mechanical support structure is shown in further detail in Figure ~\ref{fig:vibration}(a). The vacuum chamber contains two concentric copper radiation shields which are thermally anchored to the 4 and 50~K stages with flexible copper braids. The ion trap is mounted inside the cold shield near 4~K; this shield is supported by the intermediate shield near 50~K via Vespel posts. The intermediate shield is then supported directly by the room-temperature, stainless steel chamber via Macor posts. The chamber is bolted to a floating optical table. In this configuration, the ion trap is mechanically referenced via short, stiff connections to the plane of the optical table and also therefore to all beam optics, including the lens stack (mounted below the table) used for ion imaging. The low thermal conductivity of both Vespel and Macor allow the cold and intermediate shielding assemblies to be sufficiently thermally isolated from each other and from the room-temperature vacuum chamber\cite{thermal_conductivity}. Thermal connections between the trap and cold shield assembly occur via two primary paths. First, the trap die is bonded to a ceramic pin grid array (CPGA), with 100 gold pins providing both electrical and thermal connections to a PCB filter board well thermalized to the surrounding cold shield. A second thermalization path is provided by a copper bridge mounted above the trap surface as shown in Figure ~\ref{fig:chamber}(b). This component makes contact with the top surface of the alumina trap package and is also bolted to the cold shield via openings in the PCB and magnetic shielding. With the cold head at 5~K, a thermometer inside the cold shield registers a temperature of 9.9~K. This differential is due to $\sim$2.9~W of heat load flowing through the copper braids thermalizing the cold shield with conductance of 0.6~W/K. This load is dominated by $\sim$1.8~W conducted through 102 \textrm{BeCu} wires used for trap control, 0.4~W from the Vespel supports, and an estimated 0.6~W of rf dissipation.

We characterize relative lateral displacements between the optical table and the cold shield with a commercial laser Doppler vibrometer (Polytec OFV-5000 with OFV-552 fiber-optic sensor heads). We secure the sensor head to the optical table and reflect the light off a 0.5~in mirror mounted on the cold shield. We perform displacement measurements relative to the sensor head over intervals of 20~s and pass the signals through a 30~mHz high-pass filter to remove slow drifts [Fig. \ref{fig:vibration} (b)]. With the cold head powered off, we measure RMS vibrations of the cold shield to be 3.78(3)~nm. We perform spectral analysis of the vibrations by performing a discrete Fourier analysis of the 20~s time series data with 0.05~Hz resolution bandwidth [Fig. \ref{fig:vibration} (c)]. With the cold head powered on, these vibrations increase to 7.61(4)~nm. There is also a prominent peak at 1.2~Hz, the cryocooler pulsing frequency, and its harmonics. Vibrations at $\sim$30-300~Hz can be attributed to mechanical resonances of the floor and optical table which are excited when the cryocooler is on. At the achieved vibration amplitudes, we expect gate errors due to fluctuations in optical phase and intensity to be greatly reduced. 

\begin{figure}[t]
\includegraphics[width=0.48\textwidth]{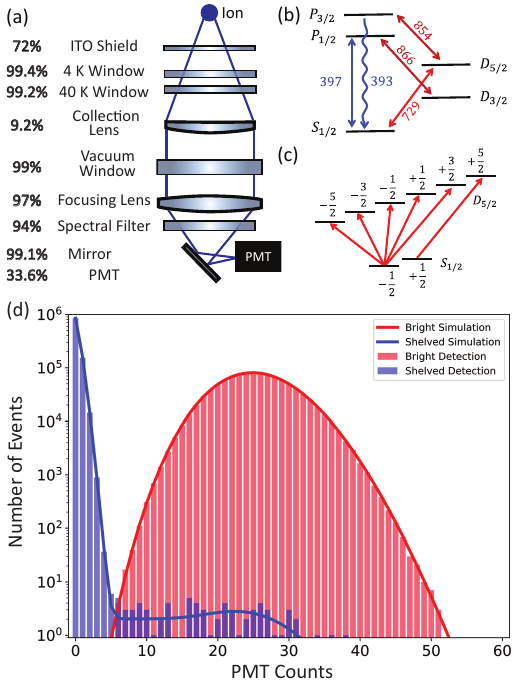}
\caption{\label{fig:det_efficiency} \textbf{(a)} Schematic of imaging stack components shown with corresponding transmission at 397~nm.  \textbf{(b)} Energy level diagram of $^{40}$Ca$^{+}$ with relevant transition wavelengths listed in ~nm. \textbf{(c)} Shelving transitions used to initialize an ion in the dark state for detection fidelity experiments.  \textbf{(d)} Fluorescence histograms for $N = 1.02\times10^6$ trials each of bright and dark using a detection interval of 50~$\mu$s. Solid lines represent a detection model (discussed in the main text). For a bright/dark discrimination threshold of 5 counts, the detection fidelity is 99.9963(4)\%, in good agreement with our modeled fidelity of 99.9969(4)\%.}
\end{figure}

State determination for trapped ions usually relies on collecting state-dependent fluorescence. Traditionally, the fluorescence is collected by an objective lens outside of the vacuum chamber and is directed to a camera or photomultiplier tube (PMT) for photon counting; with $^{40}\mathrm{Ca}^{+}$, detection error rates as low as $0.9\times10^{-4}$ in detection intervals as short as 145~$\mu$s\cite{Myerson_detect,burrell_scalable_2010} have been achieved. The photon-detection efficiency of trapped-ion systems is often limited by the light-collection solid angle of the imaging objective. Recent efforts to improve readout fidelity and reduce detection durations focus on increasing the solid angle of light collection, either by moving an objective lens into the vacuum chamber\cite{carter_ion_2024, Clark_detect} or by integrating micro-fabricated optics and sensors directly into the trap structure\cite{true_merrill_demonstration_2011, todaro_state_2021, reens_high-fidelity_2022, knollmann_collection_2025}. In our cryogenic system, we employ an in-vacuum objective with a working distance of 20.5~mm to achieve a numerical-aperture of 0.6, allowing us to collect 9.2\% of the light scattered by the ion (10\% from the solid angle with 91.7\% transmission through the lens coating). The lens is mounted on a piezo-driven hexapod [see Fig. {\ref{fig:chamber}}(a)], which allows us to remotely control the position and orientation of the optic with six degrees of freedom. Taking into account the losses from all optical interfaces in the imaging system and the quantum efficiency of the PMT, we predict an overall detection efficiency of 1.96\% [Fig. {\ref{fig:det_efficiency}}(a)].


To determine the true detection efficiency, we perform single-photon detection experiments with a trapped $^{40}\textrm{Ca}^+$ ion, similar to those described in Refs. \onlinecite{Clark_detect, Streed_2017}. Relevant transition frequencies are shown in Figure \ref{fig:det_efficiency}(b). First, the ion is Doppler cooled for 5~$\mu$s with 397 and 866~nm beams. Then, the 866~nm re-pump beam is turned off for 5~$\mu$s, allowing the atom to decay to the $D_{3/2}$ level (lifetime $\sim$1~s). The 397~nm beam is turned off and a 5~$\mu$s delay is inserted to ensure the 397~nm beam is fully extinguished. Finally, the 866~nm beam is turned on to pump population into the $P_{1/2}$ level (lifetime $\sim$7~ns). From here, the atom quickly decays to the $S_{1/2}$ level, generating a single 397~nm photon. This procedure is repeated for $10^5$ experimental trials, with the PMT detecting a photon on 1770 occasions, yielding a detection efficiency of 1.77(4)\% in good agreement with the predicted value.

Higher collection efficiency allows us to perform faster detection, leading to reduced measurement errors from spontaneous decay. To determine the detection fidelity of our system, we perform experiments to compare fluorescence histograms measured from ions prepared in a "bright" state ($S_{1/2}$ level) against those measured from ions prepared in a "dark" state ($D_{5/2}$ level). We infer the ion state from photon-count thresholding. For a specified count threshold, above-threshold events in the dark histogram contribute to false-positive errors ($\epsilon_{D}$) and below-threshold events in the bright histogram contribute to false-negative errors ($\epsilon_{B}$). The average error is calculated as $\epsilon = (\epsilon_{D}+\epsilon_{B})/2$. To determine the detection fidelity experimentally, we count the photons detected when an ion is prepared in the bright state or in the dark state during a 50$~\mu$s duration for $N = 1.02\times10^6$ trials each of bright and dark. To reduce the contributions of additional state-preparation errors to this detection fidelity estimate, we use a sequence of six laser pulses of 729~nm light to shelve the electronic population from the $S_{1/2}$ level to several states in the $D_{5/2}$ level for dark-state preparations, as shown in Fig. {\ref{fig:det_efficiency}}(c). The resulting detection histograms are plotted in Fig. {\ref{fig:det_efficiency}}(d). With a threshold of 5 or more photons used to indicate a bright ion, we calculate an average detection error of 3.7(4)$\times10^{-5}$. Note that the true detection fidelity of the ground-state qubit will be reduced from this value as an imperfect 729~nm transfer pulse to the optical qubit is first required. Techniques such as rapid adiabatic passage can be used to increase the fidelity of this transfer pulse\cite{Wunderlich20072007}.

We model this behavior assuming Poissonian statistics as in Ref. \onlinecite{det_model}. For a measured mean bright count of $\bar{n}_B$, we model the bright state probability distribution function as
\begin{equation}
    p_B(n) = P(n,\bar{n}_B)
\end{equation}
where $P(n,\bar{n}_B)$ is a Poisson distribution with mean $\bar{n}_B$. For the dark state probability distribution, we must also include the effect of spontaneous decay during detection. As we are using a muli-pulse shelving scheme, an additional delay, $t_{delay}$, of 20~$\mu$s is inserted in the model to account for decay to the $S_{1/2}$ level not addressed by the last shelving pulse. For a measured mean dark count of $\bar{n}_D$, we model the dark state probability distribution function as
\begin{multline}
p_D(n) = e^\frac{-(t_{det}+t_{delay})}{\tau} \times P(n,\bar{n}_D)\\
+(1-e^\frac{-t_{delay}}{\tau}) \times P(n,\bar{n}_B)\\
+e^\frac{-t_{delay}}{\tau} \times (1-e^\frac{-t_{det}}{\tau}) \times \frac{\Gamma(\bar{n}_B,n+1)-\Gamma(\bar{n}_D,n+1)}{\bar{n}_B-\bar{n}_D}
\end{multline}
where $\tau$ is the lifetime of the $D_{5/2}$ level ($\sim1~s$), $P(n,\bar{n}_D)$ is a Poisson distribution with mean $\bar{n}_D$, and $\Gamma(\bar{n}_D,n+1)$ is the incomplete gamma function. The first term represents background PMT counts when the ion remains shelved in the $D_{5/2}$ level; the second term represents fluorescence from ions that decay to the $S_{1/2}$ level before detection; the third term includes fluorescence from ions that decay to the $S_{1/2}$ level at some time during the detection interval\cite{det_model}. We use these probability distributions to simulate detection histograms with a detection interval, $t_{det}$, of 50~$\mu$s and mean bright and dark counts, $n_D$ and $n_B$, of 0.18 and 25.37, respectively [Fig. \ref{fig:det_efficiency}(b)]. From this, we model an average detection error of $3.1(4)\times10^{-5}$, in good agreement with the measured error.

\begin{figure}[t]
\includegraphics[width=0.48\textwidth]{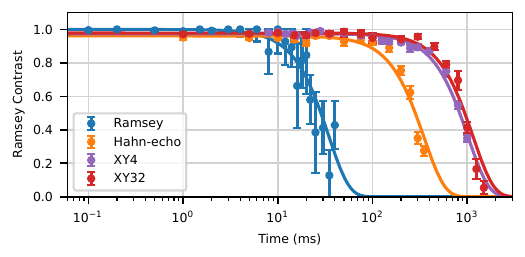}
\caption{\label{fig:coherence} Population contrast as a function of delay time for Ramsey experiments using various dynamical decoupling  schemes with rf pulses. Error bars represent the contrast fit error. Solid lines represent a Gaussian fit to the data.}
\end{figure}

Fluctuations of the magnetic field during quantum operations can lead to phase decoherence of magnetic-field sensitive qubit states. In Refs. \onlinecite{Clark_2021} and \onlinecite{Hilder_2022}, magnetic field noise was the dominant noise source limiting operation fidelity. In compact room-temperature ion-trapping systems, external magnetic shielding has been employed to reduce magnetic field fluctuations near the ions and to improve the coherence of magnetic-field-sensitive qubits to greater than 1~s\cite{long_lived_zeeman}. For a large cryogenic system, enclosing the entire vacuum chamber in multiple layers of magnetic shielding is far more difficult and costly. Instead, we designed and installed multi-layered magnetic shielding inside the vacuum chamber [Fig. {\ref{fig:chamber}}(a) (5)]. Two layers of mu-metal are mounted to the inner surface of the vacuum chamber and thermalized to room temperature. Cryo-Netic\cite{cryonetic} material lines the interior of the cold shield in both radial and axial directions with openings to accommodate cross-trap laser beams, fluorescence collection, and neutral calcium flux. Any openings can degrade the shielding performance\cite{Yashchuk_Lee_Paperno_2013}, and we mitigate this effect for the largest (imaging) aperture (under the trap) by including additional shielding layers around the in-vacuum objective and hexapod. Using COMSOL, we optimize the magnetic shielding geometry and predict a shielding factor in the central trapping region of greater than $10^{3}$ for fields applied vertically and greater than $10^{4}$ for fields applied laterally. 

To determine the efficacy of our magnetic shielding, we characterize the Ramsey coherence of the $S_{1/2}$ ground-state qubit with a magnetic sensitivity of 2.8~MHz/G. Fig. {\ref{fig:coherence}} shows the Ramsey contrast as the delay between $\pi$/2-pulses is varied. The data exhibit a Gaussian decay $C(t)=e^{(-t^2/2\tau_d^2)}$ with $\tau_d$ coherence times of  24(2)~ms without a Hahn echo and 0.25(1)~s with the echo\cite{Hahn_1950}. This is two orders of magnitude longer than the 3.8(4)~ms Ramsey Hahn-echo coherence time measured in a similar cryogenic system in our lab which does not contain magnetic shielding.



Dynamical decoupling is a powerful tool that can extend the coherence time of a qubit beyond the usual limits\cite{morong_2023, Valahu_2022}. The ability to drive dynamical decoupling pulses simultaneously on many ions at disparate locations would be advantageous for many architectures\cite{Yeh_2023, Martínez-Lahuerta_2024}.
To this end, we design a critically coupled coil resonant near the $S_{1/2}$ ground state qubit transition frequency of 21.59~MHz with a quality factor of 20. Once installed within the magnetic shielding environment, the quality factor decreases to 5, as the nearby electrically lossy high-permittivity material couples to the circuit. Despite the ensuing reduction in the rf magnetic field, we can perform qubit rotations with a $\pi$-time of 40~$\mu$s when driving the coil with 2~W. To test the performance of dynamical decoupling using the rf coil, we perform additional Ramsey coherence experiments with XY4 and XY32 (eight cycles of XY4) dynamical decoupling sequences\cite{MAUDSLEY1986} (Fig. {\ref{fig:coherence}}). We find the $\tau_d$ coherence times are extended to 0.72(2)~s and 0.81(3)~s with 4 and 32 re-focusing pulses, respectively. 

If we wish to perform dynamical decoupling pulses simultaneously at disparate trap locations, we require both a uniform biasing magnetic field and a uniform coil drive field. Initial COMSOL simulations indicate that the magnetic field produced by the coil should vary by less than 1\% over the surface of the trap. Due to the presence of magnetic shielding, we require the magnets establishing our bias field to be installed within the inner shielding layer [Fig. {\ref{fig:chamber}}(b) (9)]. Limited by available space within the cold shield, we do not achieve an optimally uniform magnet configuration. We measure a magnetic field gradient of approximately 35~mT/m, such that $S_{1/2}$ ground state Zeeman splittings differ by 0.98~kHz/$\mu$m. In our current configuration, the resulting pulse frequency and amplitude errors prevent us from performing global dynamical decoupling. In the future, improvements to the magnet configuration can reduce this non-uniformity, and some multi-pulse dynamical decoupling sequences can be implemented which are more robust to pulse amplitude errors\cite{Shappert_2013}. 


In this paper, we have presented the design and characterization of a cryogenic vacuum chamber designed to reduce error rates in trapped-ion experiments. We measure RMS vibrations of the cold shield to be 7.61(4)~nm. Our experimental photon detection efficiency is 1.77\%, allowing for fast readout of ion states. We achieve single-shot measurement fidelity of 99.9963(4)\% with a detection interval of 50~$\mu$s. We also evaluate our magnetic shielding performance by measuring Ramsey coherence times of a magnetic field sensitive qubit to be 24(2)~ms, which extends to 0.25(1)~s with a single Hahn-echo re-focusing pulse. Finally, we utilize a radio frequency antenna to demonstrate XY4 and XY32 dynamical decoupling sequences which extend the coherence of our ion to 0.72(2)~s and 0.81(3)~s, respectively. We expect these improvements to our chamber design to greatly impact our achievable algorithmic fidelity. 

This work was done in collaboration with Los Alamos National Laboratory. The authors have no conflicts to disclose. The data that support the findings of this study are available from the corresponding author upon reasonable request. 

\nocite{*}
\bibliography{main}

\end{document}